\documentstyle[prl,aps,epsf]{revtex}

\newcommand{\be}{\begin{equation}}
\newcommand{\ee}{\end{equation}}
\newcommand{\bea}{\begin{eqnarray}}
\newcommand{\eea}{\end{eqnarray}}
\newcommand{\ba}{\begin{array}}
\newcommand{\ea}{\end{array}}

\draft

\begin{document}
\twocolumn[
\hsize\textwidth\columnwidth\hsize\csname@twocolumnfalse\endcsname

\title{Intermediate-Coupling Theory for the Spectral Weight of Spin Polaron}

\author{Min-Fong Yang$^{\dagger}$ }

\vspace{0.2cm}
\address{Department of General  Programs,\\
Chang Gung College of Medicine and Technology,\\
Kweishan, Taoyuan, Taiwan, R.O.C.}

\date{\today}

\maketitle

\begin{abstract}
Within the intermediate-coupling theory, the quasiparticle weight $Z$ of one
hole injected in the undoped antiferromagnetic ground state is studied.  We
find that, for the hole located at the quasiparticle band minimun with momentum
${\bf k}_0 = ( \pm \frac{\pi}{2}, \pm \frac{\pi}{2} )$, $Z$ is finite. By
comparing the results obtained by the self-consistent Born approximation, we
show that the intermediate-coupling theory for $Z$ is appropriate only when
$J/t \stackrel{>}{\sim} 1.6$.
Finally, the reason why this approach fails in the small-$J$ case will also be
clarified.
\end{abstract}



]


The problem for the motion of a hole in a 2D antiferromagnet (AF)  has received
significant attention \cite{review} especially since the discovery of the
copper oxide superconductors, where superconductivity arises from the doping of
holes in an antiferromagnetic insulator.
The AF with one hole is also a highly non-trivial
correlated electronic system, and is therefore of fundamental
interest from a theoretical point of view.

Intuitively one would expect that the presence of a hole in an AF leads to a
distortion of the underlying spin configuration, in a similar way as a
conduction electron in a polar crystal causes a deformation of the lattice.
Indeed, under the assumptions that the AF is not completely destroied by doping
a hole and that the low-energy excitations of the spin background are spin
waves, one arrives
at  the following effective Hamiltonian \cite{SVR,KLR}, $\tilde{H}$, which is
reminiscent of Fr\"ohlich's polaron Hamiltonian \cite{Fro} :
\bea
\tilde{H} & = & H_t + H_J ,
\nonumber \\
H_J & = &   \sum_{\bf q} \omega_{\bf q} b_{\bf q}^\dagger b_{\bf q},
\label{tJ}  \\
H_t & = &  \frac{\alpha \omega}{\sqrt{N}}
  \sum_{\bf k,q} ( M_{\bf q}({\bf k}) f_{\bf k-q}^\dagger f_{\bf k}
                   b_{\bf q}^\dagger + h.c. ) \nonumber
\eea
Here $b_{\bf q}$ and $f_{\bf k}$ are the annihilation operators of the hole
and the spin wave.
$\omega_{\bf q} =\omega \nu_{\bf q}$ is
the spin wave excitation spectrum, where $\omega=Jz/2$,
$\nu_{\bf q}=\sqrt{1-\gamma_{\bf  q}^2}$, and
$\gamma_{\bf q}=\sum_{\bf d} e^{i {\bf q \cdot d}}/z$
with ${\bf d}$ the unit vectors to nearest neighbors and
$z$ being the coordination number ( $z=4$ for a  2D square lattice ).
$\alpha = 2t/J$ is the dimensionless coupling parameter ( therefore, the
small-$J$ limit means the strong-coupling limit of $\tilde{H}$ ),
$N$ is the number of the lattice sites, and
$M_{\bf q}({\bf k}) = \hbox{cosh} \theta_{\bf q} \gamma_{\bf k-q} +
\hbox{sinh} \theta_{\bf q} \gamma_{\bf k}$
is the coupling function between  the hole and the spin wave,
where $\hbox{cosh} \theta_{\bf q}=[(1+\nu_{\bf q})/(2 \nu_{\bf q})]^{1/2}$ and
$\hbox{sinh} \theta_{\bf q}
= -{\rm sgn} ( \gamma_{\bf q} ) [(1-\nu_{\bf q})/(2 \nu_{\bf q})]^{1/2}$
are the Bogoliubov transformation coefficients.
Based on this similarity \cite{note1}, a hole in an AF may be viewed as a
`` spin polaron ", i.e., a hole dressed by a cloud of virtual spin-wave
excitations of the antiferromagnetic spin background.  From this observation,
the intermediate-coupling treatment \cite{LLP} of the Fr\"ohlich polaron
problem is recently applied to the present spin-polaron problem by Barentzen
\cite{Bar}. It is found that : (1) the intermediate-coupling results for the
quasi-particle energy $E( \bf k )$ is in agreement with the dispersion curve
obtained by means of a Green function Monte Carlo method \cite{DNB}; (2) the
result for the bandwidth $W$ is quite good for
weak coupling ( $J/t > 3$ ), and is still reasonably good in the intermediate
range
( $0.4 \stackrel{<}{\sim} J/t \leq 3$ ), where the deviation from the values
obtained by the self-consistent Born approximation ( SCBA ) \cite{MH} was
about 10 - 20 \%. Thus the intermediate-coupling theory may be appropriate for
$J/t \stackrel{>}{\sim} 0.4$.

One of the most controversial issues in the spin-polaron problem is whether a
hole injected in the undoped ground state behaves like a quasiparticle
\cite{And}, or,
equivalently, whether the quasiparticle weight ( or the wavefunction
renormalization constant ) $Z$ at the Fermi surface is finite under the
dressing by the spin-wave excitations.
There have already been many studies along this line
(see \onlinecite{review} for further references).
However, most of the previous studies have involved numerical calculations
{\it on small clusters} ( even
the studies using the SCBA have to solve Dyson's equation numerically
for small clusters ).
Although numerical calculations on clusters
show that the hole has a finite quasiparticle weight,
there is still some uncertainty as to
whether the quasiparticle weight vanishes or not
in the thermodynamic limit \cite{QPW}.

In this report, we will study the quasiparticle weight within the
intermediate-coupling theory, in which we can freely take the thermodynamic
limit. Since, when a single
hole is doped, the hole will locate at the quasiparticle band minimun with
momentum
${\bf k}_0 = ( \pm \frac{\pi}{2}, \pm \frac{\pi}{2} )$ \cite{review}, we will
confine ourselves to
the spectral weight of the hole at momentum ${\bf k}_0$.
We find that  the deviation from the
results of Refs.~\cite{MH} is below 20 \%
only when $J/t \stackrel{>}{\sim} 1.6$. Thus
the range of validity for $Z$ is {\it smaller} than that
for $W$. Moreover, we will point out that, although this approach is not
plagued with the finite-size effect, due to its mean-field nature, the infrared
behavior
of the present system may not be correctly described by this method {\it even
qualitatively} ! This
may be the reason why the intermediate-coupling theory for the spin-polaron
problem is not as successful as that for the lattice-polaron case.

In order to compare with the results obtained by SCBA \cite{MH},
rather than starting from the Hamiltonian used in Barentzen's paper ( Eq.(11)
of Ref.~\cite{Bar}, which is denoted as $H_{Bar}$ in the present report ), we
take Eq.(\ref{tJ})
as our starting point, where the Bogoliubov transformation has been taken,
such that the unperturbated ground state is the vacuum state for the
spinless fermion operators and the {\it quantum} N\'eel state, $|0 \rangle$,
with respect to the spin-wave operators $b_{\bf q}$.
The relation between $\tilde{H}$ and $H_{Bar}$ is simply
\be
\tilde{H} = V^\dagger H_{Bar} V,
\ee
where $V$ denotes the unitary operator of the Bogoliubov transformation.

Following the procedure in Ref.~\cite{Bar}, one first makes
a change of coordinates to the rest frame of the moving hole by the unitary
operator ( i.e., the so-called Jost transformation ) \cite{note2}
\be
U = \sum_{\bf k, p}
( \frac{1}{N} \sum_i T_i e^{-i {\bf p} \cdot  {\bf R}_i }  )
f_{\bf k}^\dagger f_{\bf k-p}  ,
\label{Utrans}
\ee
where the translation operators for the bosons are defined by
\be
T_i = \hbox{exp} ( -i {\bf R}_i \cdot
           \sum_{\bf q} {\bf q} b_{\bf q}^\dagger b_{\bf q} ).
\label{Ti}
\ee
Then, in order to further diagonalize all terms linear in the boson operators
of the
transformed Hamiltonian, $U^\dagger \tilde{H} U$, a displacement transformation
$\tilde{W}$ is employed:
\bea
\tilde{W} &=&  \sum_{\bf k} W_{\bf k} f_{\bf k}^\dagger f_{\bf k}  ,
\label{Wtrans} \\
W_{\bf k} &=& \hbox{exp} [ \frac{1}{\sqrt{N}}
           \sum_{\bf q} \lambda_{\bf q}({\bf k})
                      ( b_{\bf q} - b_{\bf q}^\dagger ) ],
\label{Wk}
\eea
where the unknown parameters $\lambda_{\bf q}({\bf k})$ are determined by the
mean-field equations via the variational principle, i.e.,
\be
\frac{ \delta E({\bf k}) }{ \delta \lambda_{\bf q}({\bf k}) }  = 0,
\ee
with the ground-state energy $E({\bf k})$ defined by the expectation value
\be
E({\bf k}) = \langle f_{\bf k}^\dagger, 0 |
               \tilde{W}^\dagger U^\dagger \tilde{H} U \tilde{W}
             | f_{\bf k}^\dagger, 0 \rangle .
\label{Ek}
\ee
It can be shown that, for a single hole, the Jost transformation $U$ and the
Bogoliubov transformation $V$ commute each other, then
\bea
& & \langle f_{\bf k}^\dagger, 0 |
               \tilde{W}^\dagger U^\dagger \tilde{H} U \tilde{W}
             | f_{\bf k}^\dagger, 0 \rangle  \nonumber \\
&=& \langle f_{\bf k}^\dagger, 0 |
          \tilde{W}^\dagger U^\dagger V^\dagger H_{Bar} V U \tilde{W}
             | f_{\bf k}^\dagger, 0 \rangle    \\
&=& \langle f_{\bf k}^\dagger, 0 |
          \tilde{W}^\dagger V^\dagger U^\dagger H_{Bar} U V \tilde{W}
             | f_{\bf k}^\dagger, 0 \rangle , \nonumber
\eea
which is just the expression of the expectation value obtained in
Ref.~\cite{Bar}.
Thus the variational calculations in our case is completely the same as those
in Ref.~\cite{Bar}. Therefore, the mean-field equations lead to the following
self-consistent equations: ( see Eqs.(57)-(58) of Ref.~\cite{Bar} ),
\bea
\Omega ({\bf k})
  &=& \frac{\alpha}{N} F^2({\bf k})
      \sum_{\bf q}
      \frac{M^2_{\bf q}({\bf k})}{\nu_{\bf q} + 2 \alpha \Omega({\bf k})} ,
\label{SC1} \\
\hbox{ln} F({\bf k})
  &=& - \frac{\alpha^2}{N} F^2({\bf k})
      \sum_{\bf q}
      \frac{M^2_{\bf q}({\bf k})}{(\nu_{\bf q} + 2 \alpha \Omega({\bf k}))^2},
\label{SC2}
\eea
where $F({\bf k})$ and $\Omega({\bf k})$ are defined in terms of
$\lambda_{\bf q} ({\bf k})$ as
\bea
F({\bf k})
 & = & \hbox{exp} (
            - \frac{1}{N} \sum_{\bf q} \lambda^2_{\bf q}({\bf k})
           ) ,
\label{Fk}     \\
\Omega({\bf k})
 & = & \frac{1}{N} F({\bf k})
       \sum_{\bf q} \lambda_{\bf q}({\bf k}) M_{\bf q} ({\bf k}).
\eea
Then the ground-state energy can be written in terms of $F({\bf k})$ and
$\Omega({\bf k})$ :
\be
E({\bf k}) = - \alpha \omega \Omega({\bf k}) [ 1 - 2 \hbox{ln} F({\bf k}) ].
\ee

Now we turn to the calculation of the spectral weight. The spectral weight
$Z_{\bf k}$ of the hole at momentum ${\bf k}$ is defined as
\be
Z_{\bf k} =
| \langle  \phi^{(0)}_{\bf k} | \phi_{\bf k} \rangle |^2 ,
\label{QPW}
\ee
where $| \phi^{(0)}_{\bf k} \rangle = | f_{\bf k}^\dagger, 0 \rangle$ is the
ground-state eigenvector of the unperturbated Hamiltonian, $H_J$, within the
one-hole subspace;
while $| \phi_{\bf k} \rangle = U \tilde{W} | f_{\bf k}^\dagger, 0 \rangle$ is
the corresponding variational ground-state eigenvector of the full
Hamiltonian, $\tilde{H}$. By Eqs.(\ref{Utrans})-(\ref{Wk}),
\bea
\langle \phi^{(0)}_{\bf k} | \phi_{\bf k} \rangle
&=& \langle 0 | \frac{1}{N} \sum_i T_i W_{\bf k} | 0 \rangle
                      \nonumber \\
&=& \langle 0 | W_{\bf k} | 0 \rangle
\label{phi}                      \\
&=& \hbox{exp} (
            - \frac{1}{2N} \sum_{\bf q} \lambda^2_{\bf q}({\bf k})
           ) ,  \nonumber
\eea
where $T_i^\dagger |0 \rangle = |0 \rangle $ ( because
$b_{\bf q} |0 \rangle =0$ ) and the Baker-Hausdorff formula is used in the
last line of the derivation . From Eqs.(\ref{QPW}), (\ref{phi}), and
(\ref{Fk}), one obtains \bea
Z_{\bf k}
&=& \hbox{exp} (
            - \frac{1}{N} \sum_{\bf q} \lambda^2_{\bf q}({\bf k})
           ) , \nonumber \\
 &=& F({\bf k}).
\eea
Thus, by solving $F({\bf k})$ and $\Omega({\bf k})$ via the self-consistent
equations, Eqs.(\ref{SC1}) and (\ref{SC2}), we get $Z_{{\bf k}}$ at the same
time.

From now on, we confine ourselves to the case of
${\bf k} = {\bf k}_0$. After numerically solving
the self-consistent equations,
we get $Z = Z_{{\bf k}_0}$ as a function of $J/t$.  The result is shown in
Fig.$1$. If we compare our results ( solid line ) with those obtained by SCBA
\cite{MH} ( open circles ), we realize that our spectral weight is fairly
accurate for $J/t > 2.5$ ( i.e., the agreement is within 10 \% ). However,
 the deviation from the
results of Refs.~\cite{MH} is below 20 \% only when
$J/t \stackrel{>}{\sim} 1.6$.  Thus the range of validity for $Z$ is
{\it smaller} than that for $W$.

\begin{figure}
\centerline{\epsfysize=7.5cm
\epsfbox{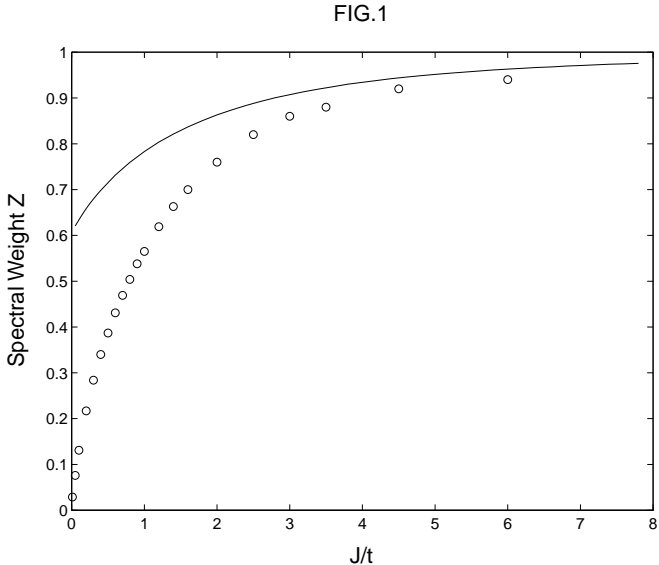}}
\caption{
The spectral weight $Z$ for one hole
located at the quasiparticle band minimun with momentum
${\bf k}_0 = ( \pm \frac{\pi}{2}, \pm \frac{\pi}{2} )$ as a function of
$J/t$.  The results of the intermediate-coupling theory are shown as the solid
line, and open circles represent those obtained by SCBA ~\protect\cite{MH}.
}
\end{figure}

In the following, we will clarify the reason why the intermediate-coupling
theory for the spin-polaron problem is not as successful as that for the
lattice-polaron case.
Notice that,
after restricting to the subspace with one hole at momentum ${\bf k}$,
as shown in Eqs.(45c) and (47) of Ref.\cite{Bar},
one arrives at the effective Hamiltonian for the boson operators \cite{note3}
\be
{\cal H}_{\bf k}
\simeq \hbox{constant} +
\sum_{\bf q} \epsilon_{\bf q} ({\bf k}) b_{\bf q}^\dagger b_{\bf q},
\ee
where
$\epsilon_{\bf q}({\bf k}) = \nu_{\bf q} + 2 \alpha \Omega({\bf k})$ can be
considered as the
{\it renormalized} energy spectrum of the spin waves in unit of $\omega$ in the
rest frame of the hole at momentum ${\bf k}$.
Hence, for ${\bf q} \to 0$, the renormalized energy spectrum
$\epsilon_{\bf q}({\bf k}) \to 2 \alpha \Omega({\bf k})$, which is nonvanishing
for a given ${\bf k}$; while the `` bare " one,
$\nu_{\bf q} \propto |{\bf q}| \to 0$. That is, a {\it finite gap},
$2 \alpha \Omega({\bf k})$, is
introduced into the excitation spectrum of the spin waves by this variational
approach, even though the original excitation spectrum is gapless due to the
Goldstone theorem \cite{Gold} ! Thus the long-wavelength ( infrared )
properties of
the system is altered {\it qualitatively}. This qualitative change in
the excitation spectrum is not reasonable, because the correction to the spin
wave energies by a single hole in a macroscopic AF background should be
proportional to $1/N$, which would be negligible in the thermodynamic limit !
In the lattice-polaron case, although there is still a nonvanishing
correction to the phonon energies \cite{LLP}, there is {\it no} qualitative
change in the phonon energies by this approach, because only the longitudinal
optical ( LO ) phonons are considered and the `` bare " energy spectrum of
these LO phonons can approxmately be taken as a positive constant. As claimed
by Anderson \cite{And}, whether $Z$ is zero or not greatly depends on the
infrared behavior of the system. Since the infrared behavior may not be
faithfully described by this variational approach, it is reasonable that the
intermediate-coupling theory may not predict the accurate value of $Z$ in
the spin-polaron case.

For a further support of the above arguements, we present the results of
the ``induced spin gap",
$2 \alpha \Omega({\bf k}_0)$, as a function of $J/t$ in Fig.2. We find that, as
$J/t$ decreases, the ``induced spin gap" increases and
invalidates the variational approach.

\begin{figure}
\centerline{\epsfysize=7.5cm
\epsfbox{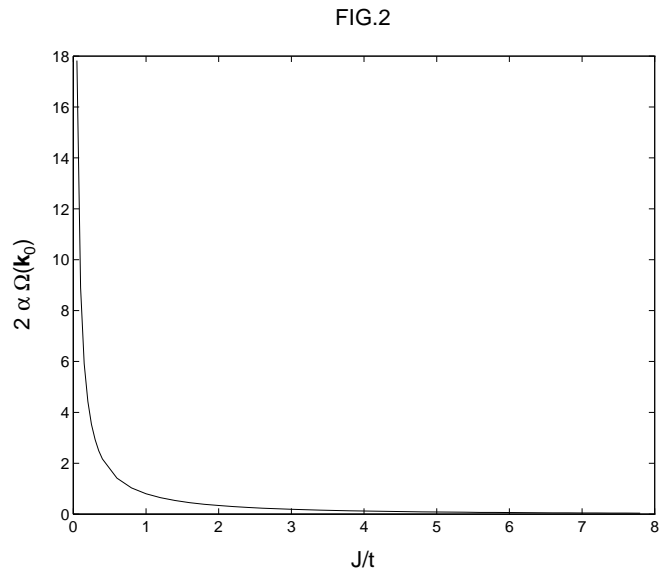}}
\caption{
The `` induced spin gap ", $2 \alpha \Omega({\bf k}_0)$, as a function
of $J/t$.
}
\end{figure}

In conclusion, within the intermediate-coupling theory, we show that
the quasiparticle weight $Z$ is finite, and our results agree with those
obtained by the self-consistent Born approximation \cite{MH}
when $J/t \stackrel{>}{\sim} 1.6$. Because of the failure to describe
correctly the
infrared behavior of the system, this approach is not suitable for the study of
the spectral weight of the spin-polaron case, especially when $J/t$ is small.

\noindent{\bf Acknowledgment:} The author thanks Prof. T. K. Lee and Dr. M. C.
Chang for their critical reading of the manuscript.

$\dagger$ E-mail address: mfyang@phys.nthu.edu.tw

%
%
%
%
%


\begin{references}
\bibitem{note2}  In order to link to the treatment in Ref.~\protect\cite{Bar},
        we will follow his notations as closely as possible.
\bibitem{note3}  The off-diagonal terms and the terms of higher than quadratic
        order in the boson operators are omitted, which are irrelevant to our
        present discussions.
\end{references}
\end{document}